\documentclass[journal]{IEEEtran}
\usepackage{amsfonts}
\usepackage{amssymb}
\usepackage{amsmath}
\usepackage[dvips]{graphicx,psfrag,overpic,color}
\usepackage{cite}
\usepackage{array}
\usepackage[normalem]{ulem}
\usepackage{bm}

\begin{document}

\title{Lorenz System Parameter Determination and Application to
Break the Security of Two-channel Chaotic Cryptosystems}

\author{A.~B.~Orue, G.~Alvarez,  M.~Romera, G.~Pastor, F.~Montoya\thanks{A.~B.~Orue,
G.~Alvarez,  M.~Romera, G.~Pastor and F.~Montoya are with the
Instituto de F\'{\i}sica Aplicada, Consejo Superior de
Investigaciones Cient\'{\i}ficas, Serrano 144, 28006 Madrid, Spain
(Email: gonzalo@iec.csic.es).} and Shujun Li\thanks{Shujun Li is
with the Department of Electronic and Information Engineering, The
Hong Kong Polytechnic University, Hung Hom, Kowloon, Hong Kong SAR,
China.}}


\maketitle

\begin{abstract}
This paper describes how to determine the parameter values of the
chaotic Lorenz system used in a two-channel cryptosystem. The
geometrical properties of the Lorenz system are used firstly to
reduce the parameter search space, then the parameters are exactly
determined, directly from the ciphertext, through the minimization
of the average jamming noise power created by the encryption
process.
\end{abstract}

\sloppy

\section{Introduction}
\PARstart{I}{n} recent years, a growing number of cryptosystems
based on chaos synchronization have been proposed~\cite{yang04},
many of them fundamentally flawed by a lack of robustness and
security.

The first schemes of synchronization-related chaotic cryptography
were based on the masking of a plaintext message by a system
variable of a chaotic generator \cite{Boutayeb02, Memon03,
Bowong04}. The receiver had to synchronize with the sender to
regenerate the chaotic signal and thus recover the message. This
simple design is easily broken by elemental filtering of the
ciphertext signal \cite{Alvarez04f,Alvarez04g,Alvarez05b}.

Recently, there appeared some chaotic cryptosystems with an
enhanced plaintext concealment mechanism: the ciphertext consisted
of a complicated non-linear combination of the plaintext and a
variable of a chaotic transmitter generator, from which it was an
unattainable goal to retrieve a clean plaintext. As it was
impossible to synchronize a chaotic receiver with such ciphertext,
a second channel was used for synchronization. The synchronizing
signal was a different sender chaotic variable, that was
transmitted without modification. The same system parameters
values were used at sender and receiver
\cite{Jiang02,Wang04,Li04}.

One of these cryptosystems, proposed by Jiang~\cite{Jiang02}, made
use of the Lorenz chaotic system~\cite{Lorenz63}, that is defined
by the following equations:
\begin{align}
  \dot {x}  &= \sigma(y-x),\nonumber\\
  \dot {y}  &=\rho x - y -xz,\label{eq:sender}\\
  \dot {z}  &=xy - \beta z,\nonumber
\end{align}
where $\sigma$, $\rho$ and $\beta$ are fixed parameters.

The ciphertext $s$ was defined as
\begin{equation}
s=f_1(x,y,z)+f_2(x,y,z)\,m,\label{eq:ciphertext0}
\end{equation}
where $m$ is the plaintext.

The receiver was designed as a reduced order nonlinear observer
with a mechanism to achieve efficient partial synchronization,
under the drive of $x(t)$. It can generate two signals $y_r(t)$
and $z_r(t)$ that converge to the driver system variables $y(t)$
and $z(t)$, respectively, as $t\to\infty$.

The recovered plaintext $m^*(t)$ was retrieved with the function:
\begin{equation}
m^* =
\frac{s}{f_2(x,y_r,z_r)}-\frac{f_1(x,y_r,z_r)}{f_2(x,y_r,z_r)}.\label{eq:retrieved0}
\end{equation}

It was given an example in~\cite[\S III]{Jiang02} with the
following functions: $f_1(x,y,z)=y^2$ and $f_2(x,y,z)=1+y^2$; the
following parameter values: $\sigma=10$, $\rho=28$ and
$\beta=8/3$; and with the following initial conditions: $(x(0),\,
y(0),\, z(0))=(0,\, 0.01,\, 0.01)$ and $(y_r(0),\,
z_r(0))=(0.05,\, 0.05)$. The plaintext was a small amplitude
sinusoidal signal of 30 Hz, $m(t)=0.05\sin(2\pi30t)$. The author
claimed that this cryptosystem guarantees higher security and
privacy, showing that an error of 0.05 in the retrieval of $y_r$,
due to a poor parameter estimation, giving rise to a serious
distortion in the retrieved plaintext.

In the vast majority of chaotic cryptosystems, the security relies
on the secrecy of the system parameters, which play the role of
secret key. Hence, the determination of the system parameters is
equivalent to breaking the system. Recently, Solak~\cite{solak04}
analyzed the cryptosystem~\cite{Jiang02} and showed how an
eavesdropper could identify the value of the parameter $\rho$,
provided that it has the previous knowledge of the two other
transmitter system parameters $\beta$ and $\sigma$. Solak's
approach was based on a novel expression of the Lorenz system.
Formerly Stojanovski, Kocarev and Parlitz~\cite{Stojanovski96}
described a generic method, to reveal simultaneously all the three
parameters of a Lorenz system when one of the the variables $x(t)$
or $y(t)$ were known, that could be applied to break this
cryptosystem.

The present work describes an efficient determination method of
the only two unknown parameters $\rho$ and $\beta$ needed to build
up an intruder Lorenz system receiver, from the ciphertext alone,
without partial knowledge of any transmitter parameters. Firstly,
some geometrical properties of the Lorenz attractor are shown.
Then, advantage is taken of them to minimize, as much as possible,
the parameters search space. Finally, the unknown receiver
parameters are determined with high accuracy.

\section{The Lorenz attractor's geometrical
properties}\label{sec:geometrical}

According to \cite{Lorenz63}, the Lorenz system has three
equilibrium points. The origin is an equilibrium point for all
parameter values; for $0<\rho<1$ the origin is a globally attracting
asymptotically stable sink; for $1\leq \rho \leq \rho_H$ the origin
becomes a non-stable saddle point, giving rise to two other stable
twin equilibrium points $C^+$ and $C^-$, of coordinates
$x_{C^\pm}=\pm\sqrt{\beta (\rho -1)}$, $y_{C^\pm}=\pm\sqrt{\beta
(\rho -1)}$ and $z_{C^\pm}=\rho -1$, being $\rho_H$ a critical
value, corresponding to a Hopf bifurcation \cite{Sparrow82}, whose
value is:
\begin{equation}\label{eq:critical}
\rho_H=\frac{\sigma(\sigma+\beta+3)}{(\sigma-\beta-1)}.
\end{equation}

When $\rho$ exceeds the critical value $\rho_H$ the equilibrium
points $C^+$ and $C^-$ become non-stable saddle foci, by a Hopf
bifurcation, and the strange Lorenz attractor appears. The flow,
linearized around $C^+$ and $C^-$, has one negative real
eigenvalue and a complex conjugate pair of eigenvalues with
positive real part. As a consequence, the equilibrium points are
linearly attracting and spirally repelling.

\begin{figure}[t ]
\center
\begin{overpic}[scale=1]{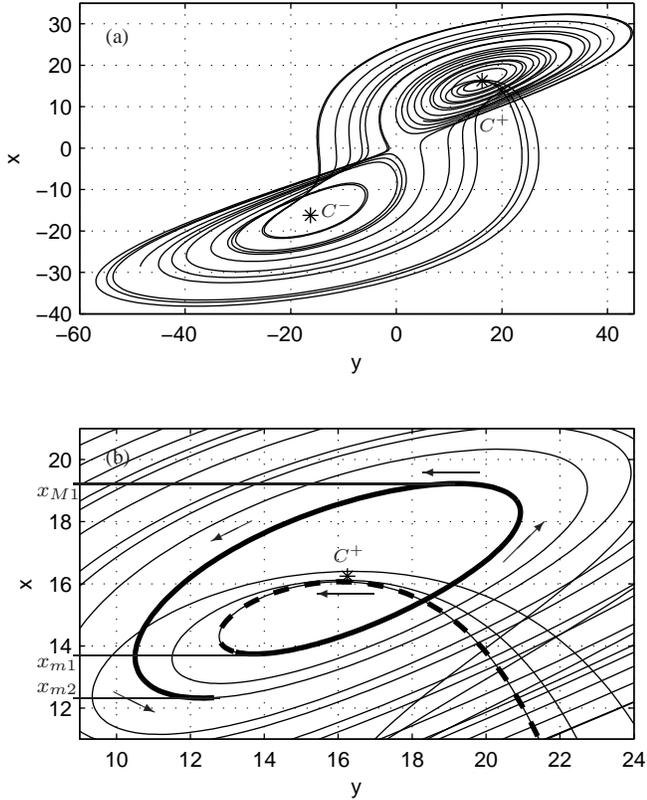}
    \put(60, 83)    {\scriptsize$C^+$}
    \put(40.5, 72.5){\scriptsize$C^-$}
    \put(42, 30)    {\scriptsize$C^+$}
    \put(47, 26)    {\vector(-1,0){7}}
    \put(32, 35)    {\vector(-2,-1){5}}
    \put(63, 30)    {\vector(1,1){5}}
    \put(60, 41)    {\vector(-1,0){7}}
    \put(15, 14)    {\vector(2,-1){5}}
    \put(10, 39.6)  {\line(1,0){47}}
    \put(10, 18.5)  {\line(1,0){23}}
    \put(10, 13.2)  {\line(1,0){18}}
    \put(5.5, 17)   {\footnotesize$x_{m1}$}
    \put(5.5, 14)   {\footnotesize$x_{m2}$}
    \put(5.5, 38.1) {\footnotesize$x_{M1}$}
    \put(14, 94)    {\footnotesize (a)}
    \put(14, 42.1)  {\footnotesize (b)}
\end{overpic}
\caption{\label{fig:atractor} Lorenz chaotic attractor; (a) $x-y$
plane projection; (b) enlarged view,  showing the incoming
trajectories portion attracted by the equilibrium point $C^+$, the
flow direction is indicated by arrows. The position of the
equilibrium points $C^+$ and $C^-$ is indicated by asterisks.}
\end{figure}

Figure~\ref{fig:atractor}(a) shows the double scroll Lorenz
attractor formed by the projection on the $x-y$ plane, in the
phase space, of a trajectory portion extending along 12 s; the
parameters are $\sigma=16$, $\rho=100$  and $\beta=8/3$.

It is a well known fact that the Lorenz attractor trajectory draws
two 3D loops, in the vicinity of the equilibrium points $C^+$ and
$C^-$, with a spiral like shape of steadily growing amplitude,
jumping from one of them to the other, at irregular intervals, in
a random like manner though actually deterministic
\cite{Lorenz63}. The trajectory may pass arbitrarily near to the
equilibrium points, but never reach them while in chaotic regime.

\begin{figure}
\center
\begin{overpic}[scale=1]{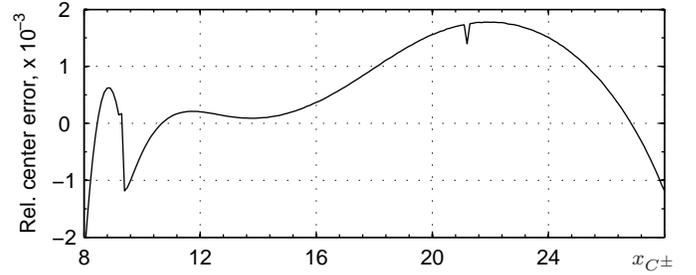}
     \put(95,0.9){\footnotesize$x_{C^\pm}$}
\end{overpic}
\caption{Equilibrium points $C^\pm$ estimation relative error,
when taking the eye center coordinate $x^*_{C^\pm}$ instead of the
true value of $x_{C^\pm}$.}\label{fig:Cerror}
\end{figure}

The geometrical properties of Lorenz system allows for a previous
reduction of the search space of the $\rho$ and $\beta$
parameters, taking advantage of the relation of them with the
coordinates $x_{C^\pm}=\pm \sqrt {\beta(\rho-1)}$  of the
equilibrium points.

Let us call attractor \textit{eyes} to the two neighborhood
regions around the equilibrium points that are not filled with the
spiral trajectory. The eye centres are the fixed points $C^+$ and
$C^-$.

The pending problem is to determine the eye centres when the inner
turns are missing, as happens in normal chaotic regime. With the
drive signal $x(t)$, we solved it by experimentally estimating the
middle point value of the trajectory maxima and minima in the
phase space projection on the $x-y$ plane. The best result was
obtained by taking into account only the regular spiral cycle
closest to the center, shown in Fig.~\ref{fig:atractor}(b) as a
thick continuous line. The $x$-coordinate of the eye center was
calculated with the following empirical formula:
\begin{equation}\label{eq:centre}
x^*_{C^\pm}=\frac{0.9~x_{m1}+0.1~x_{m2}+x_{M1}}{2},
\end{equation}
where $x_{M1}$ is the minimum of all the maxima of $|x(t)|$ spiral
trajectory, $x_{m1}$ and $x_{m2}$ are the two minima immediately
preceding and following $x_{M1}$, respectively.

As the spiral has a growing radius, it was necessary to take a
weighted mean between the two minima $x_{m1}$ and $x_{m2}$, the
optimal values of the two weights were determined experimentally.
Instead of making two computations, one around $C^+$ and another
around $C^-$, a unique computation was done on the absolute value
waveform $|x(t)|$. It should be noted that all the first maxima
after a change of sign of $x(t)$ and $y(t)$ must be discarded
because they belong to the incoming trajectory portion attracted
by the equilibrium points $C^\pm$ and do not belong to the spiral
trajectory, one of them is shown in Fig.~\ref{fig:atractor}(b) as
a thick dashed line.

The result is illustrated in Fig.~\ref{fig:Cerror}. It can be seen
that the relative value of the error, taking the eye center
coordinate $x^*_{C^\pm}$ instead of the true value of $x_{C^\pm}$,
is less than $2\times 10^{-3}$. The system parameters were varied
in the margins: $\sigma \in (9.7,37.4)$, $\rho \in (25.6, 94.8)$
and $\beta \in (2.6,8.4)$. The system initial conditions were the
same as the example of \cite[\S III]{Jiang02}; the period of
measurement was 20 s and the sampling frequency was 1200 Hz.

In this way, the search space of the unknown parameters $\beta$
and $\rho$ is reduced to a narrow margin defined as
$\beta^*(\rho^*-1) \in
\{0.996~x^{*\,2}_{C^\pm},1.004~x^{*\,2}_{C^\pm}\}$.

Applying this method to the proposed example of~\cite[\S
III]{Jiang02}, whose equilibrium point is $x_{C^\pm}=\sqrt{72}$,
the absolute determination error of $x^*_{C^\pm}$ was $7.5 \times
10^{-4}$, equivalent to a relative error of $0.0089\%$ .

\section{Breaking of the proposed encryption system}

We designed an intruder receiver based on a homogeneous driving
synchronization mechanism \cite{Pecora91} between the transmitter
drive Lorenz system and a receiver response subsystem, that was a
partial duplicate of the drive system reduced to only two
variables $y_r(t)$ and $z_r(t)$, driven by the drive variable
$x(t)$. The response system was defined by the following
equations:
\begin{align}
    \dot {y}_r  &= \rho^* x - y_r -x z_r,\nonumber\\
    \dot {z}_r  &= xy_r - \beta^* z_r. \label{eq:receiver}
\end{align}

Note that for breaking the system it is only necessary to get the
knowledge of the parameters $\rho$ and $\beta$, i.e. the parameter
$\sigma$ may be ignored and need not be determined, unlike in the
Solak method~\cite{solak04} which requires its previous knowledge,
or in the Stojanovski et all. method~\cite{Stojanovski96} which
requires the simultaneous determination of all the three unknown
parameters.

As it was shown in~\cite[$\S$III]{Pecora91}, this drive-response
configuration has two conditional Lyapunov exponents, both fairly
negative, thus leading to a very estable system. The consequence is
that, if the parameters of drive and response systems are moderately
different, the drive and response variables will be alike, though
not totally identical. This property may be exploited to search the
right parameter values looking at the retrieved plaintext and
applying an optimization procedure to find the parameters that
provide the best retrieved plaintext quality.

\begin{figure}
\begin{center}
\begin{overpic}[scale=1]{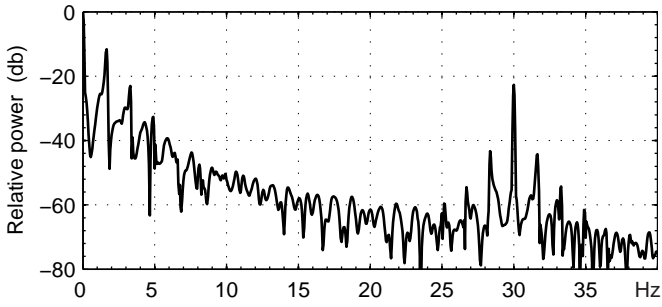}
     \put(96.5,0.8){\footnotesize \textsf{Hz}}
\end{overpic}
\caption{Logarithmic power spectrum of the retrieved plaintext
with a wrong guessing of response system
parameters.}\label{fig:spectrum}
\end{center}
\end{figure}

When the synchronizing signal is fed to the response system
described by Eq.~(\ref{eq:receiver}) and the parameters of both
systems agree, i.e. $\rho^*=\rho$ and $\beta^*=\beta$, the
variables $y$ and $y_r$ of the drive an response systems are
equal, hence the recovered text $m^*(t)$ follows the plaintext
$m(t)$ exactly; being negligible the effect of different initial
conditions after a very short transient. If the parameters of both
systems do not agree, the recovered text will consist of a noisy
distorted version of the original plaintext, growing the noise and
distortion as the mismatch between drive and response systems
parameters grows.

\subsection{Parameter determination}\label{determination}
In the particular case of the example in~\cite[$\S$III]{Jiang02},
the encryption and decryption functions were:
\begin{eqnarray}
  s&=& y^2+(1+y^2)\,m , \label{eq:ciphertext}\\
  m^*&=&\frac{s}{1+y_r^2}-\frac{y_r^2}{1+y_r^2}\,.\label{eq:retrieved}
\end{eqnarray}
Equation~(\ref{eq:retrieved}) of the recovered text can be
rewritten as:
\begin{equation}
m^* = m\,\frac{1+y^2}{1+y_r^2}+\frac{y^2-y_r^2}{1+y_r^2}.
\label{eq:retrieved2}
\end{equation}
This equation has two terms, the first one is a function of the
plaintext message $m(t)$ and the variables $y$ and $y_r$. When
$y=y_r$ the term is reduced to the undistorted plaintext, but if
$y\ne y_r$ a distortion appears. The second term is a function of
$y$ and $y_r$ and can be considered as a jamming noise.
Figure~\ref{fig:spectrum} depicts the spectrum of the recovered
text corresponding to the example, but with a wrong guessing of
the response system parameters: $\rho^*=28.01$ and
$\beta^*=2.667$. It can be seen that the spectrum has two main
frequency bands: one around the plaintext $m(t)$ frequency of
$30\,\textrm{Hz}$, that corresponds to the distorted plaintext,
and another near $0 \,\textrm{Hz}$ that corresponds to the jamming
noise. Assuming that the plaintext will always consist of an a.c.
band limited signal without d.c. component, as in the numerical
example given in~\cite{Jiang02}, it is clear from
Fig.~\ref{fig:spectrum} that the second term of
Eq.~(\ref{eq:retrieved2}) may be isolated from the first by means
of a suitable filter.

\begin{figure}
\begin{center}
\psfrag{Error power (db)}{\textsf{\footnotesize Average noise
power (db)}}
\begin{overpic}[scale=1]{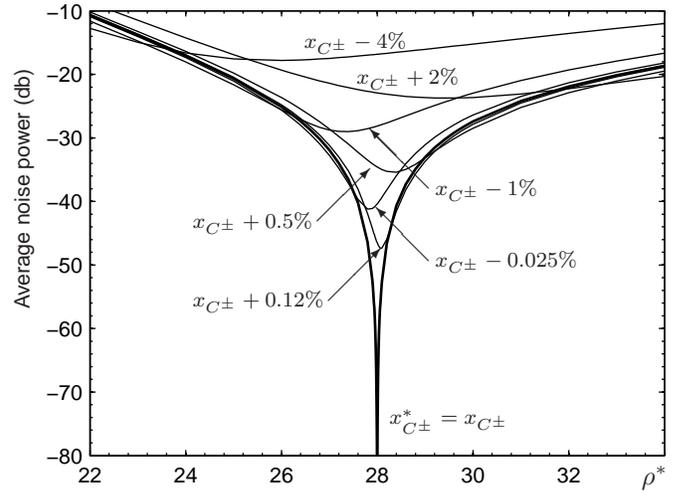}
     \put(96.5,0.8){$\rho^*$}
     \put(58, 10){\footnotesize $x^*_{C^\pm}=x_{C^\pm}$}
     \put(65, 34){\footnotesize $x_{C^\pm}-0.025 \%$}
     \put(64, 35){\vector(-1,1){8} }
     \put(65, 45){\footnotesize $x_{C^\pm} - 1 \%$}
     \put(64, 46){\vector(-1,1){9} }
     \put(28, 28){\footnotesize $x_{C^\pm} + 0.12 \%$}
     \put(48.7, 28.8){\vector(1,1){8} }
     \put(28, 40){\footnotesize $x_{C^\pm} + 0.5 \%$}
     \put(47, 41){\vector(1,1){8} }
     \put(53, 62){\footnotesize $x_{C^\pm} + 2 \%$}
     \put(45, 67.5){\footnotesize $x_{C^\pm} - 4 \%$}
\end{overpic}
\caption{Logarithmic representation of the mean of the recovered
text noise power $\overline{\varepsilon^2}$, for several values of
$x^*_{C^\pm}$.}\label{fig:noisepower}
\end{center}
\end{figure}
The most important band of the jamming noise $\varepsilon$ was
isolated by means of a finite impulse response low pass filter
with 2048 terms and a cutoff frequency of 0.2\,Hz, that suppressed
the contribution of the plaintext $m(t)$ and most of the frequency
terms generated by the modulation with the chaotic signal
$y^2(t)$. Figure\,\ref{fig:noisepower} illustrates the mean value
of the squared noise $\overline{\varepsilon^2}$, i.e. the average
noise power, as a function of $\rho^*$, with the eye center
$x^*_{C^\pm}$ as parameter, with the same transmitter system
parameters of the numerical example presented in \cite{Jiang02}
and the intruder receiver described by Eq.~(\ref{eq:receiver}).
The mean of $\varepsilon^2$ was computed along the first 20 s,
after a delay of 2 seconds, to let the initial transient finish.
It is clearly seen that the noise grows monotonically with the
mismatch between the transmitter and receiver parameters
$|\rho^*-\rho|$ and that the minimum error corresponds to the
receiver system parameter $\rho^*$ exactly matching the
transmitter system parameter $\rho$, when
$x^*_{C^\pm}=x_{C^\pm}=\sqrt{\beta(\rho -1)}=\sqrt{72}$.

The search of the correct parameter values $\beta^*$ and $\rho^*$
is carried out with the following procedure:
\begin{enumerate}
    \item Determine the approximate value of the eye center
$x^*_{C^\pm}$ as described in Section~\ref{sec:geometrical} from
the $x(t)$ waveform.
    \item Keeping the last value of $x^*_{C^\pm}$, vary the value of
$\rho^*$ until a minimum of the average noise power is reached.
    \item Keeping the last value of $\rho^*$, vary the value of eye
    center $x^*_{C^\pm}$ until a new minimum of the average noise
    power is reached.
    \item Repeat the two previous steps until a stable result of average
noise power will be reached and retain the last values of $\rho^*$
and $x^*_{C^\pm}$ as the ultimate ones.
    \item Calculate the value of $\beta^*$ as
$\beta^*=(x^*_{C^\pm})^2/(\rho^*-1)$.
\end{enumerate}

Table~\ref{tab:convergency} shows the evolution of the relative
eye center error, the relative $\rho^*$ parameter error and the
average jamming noise power. It can be seen that the procedure
converges rapidly to the exact values: $\rho^*=\rho=28$ and
$x^*_{C^\pm}=x_{C^\pm}=\sqrt{72}$.

\begin{table}[h]\caption{Evolution of the the
relative eye center error, the relative $\rho^*$ parameter error
and the average jamming noise power.}\label{tab:convergency}
\centering
\begin{tabular}{cccc}
\hline
 Step & Relative eye center error   & Relative $\rho^*$
error & Average noise\\
 & $(x^*_{C^\pm}-x_{C^\pm})/x_{C^\pm}$ &
         $(\rho^*-\rho)/\rho$   & power $\overline{\varepsilon^2}$ \\
\hline
       1  & $8.90\times 10^{-5~}$ &                &     \\
       2  & $8.90\times 10^{-5\bullet}$ & $-3.57\times 10^{-8~}$ & $5.2\times 10^{-8~}$\\
       3  & $2.72\times 10^{-8~}$ & $-3.57\times10^{-8\bullet}$ & $8.9\times10^{-12}$\\
       4  & $2.72\times 10^{-8\bullet}$ & $~~\,0~~$           & $6.5\times10^{-13}$\\
       5  & $\,0~~$            & $~~0^\bullet$           & $6.1\times10^{-13}$\\
       6  & $0^\bullet$            & $~~\,0~~$           & $6.1\times10^{-13}$\\
\hline
\multicolumn{4}{l}{$\bullet=$ old data held from the previous step} \\
\end{tabular}
\end{table}

The value of the unknown parameter $\beta^*$ was deduced from
Eq.~(\ref{eq:centre}) with the estimated values of $\rho^*$ and
$x^*_{C^\pm}$ as
$\beta^*=\frac{(x^*_{C^\pm})^2}{(\rho^*-1)}=\frac{8}{3}$.

Note that this method works as well for the general case described
by Eqs.~(\ref{eq:ciphertext0}) and (\ref{eq:retrieved0}) that have
similar structure to Eqs.~(\ref{eq:ciphertext}) and
(\ref{eq:retrieved}) which describe the special case of the
example in~\cite[$\S$III]{Jiang02}, just selected here for
experimental demonstration.

\subsection{Plaintext retrieving}

\begin{figure}
\begin{center}
\psfrag{mt}{$m^*(t)$}
\includegraphics{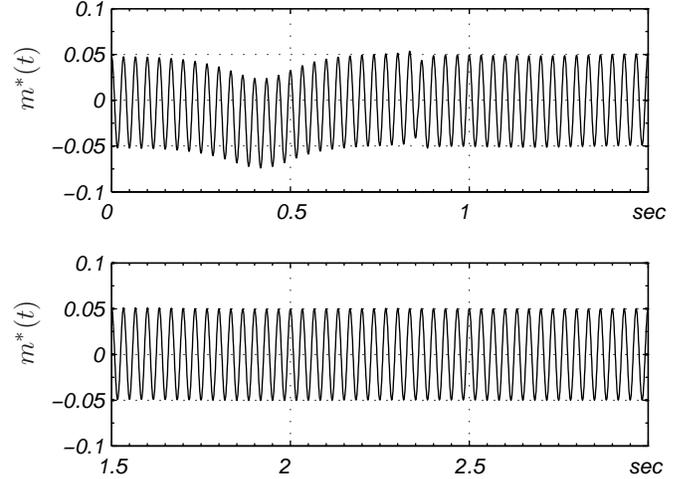}
\caption{Retrieved plaintext with the ultimate values of the
response system parameters.}\label{fig:retrieved}
\end{center}
\end{figure}

As the system parameters are equivalent to the system key, once the
exact values of $\beta^*$ and $\rho^*$ are known, the ciphertext can
be efficiently decoded by the intruder receiver defined by
Eq.~(\ref{eq:receiver}). Figure~\ref{fig:retrieved} shows the three
first seconds of the retrieved plaintext with the response system
receiver described by Eq.~(\ref{eq:receiver}), corresponding to the
ciphertext example of~\cite[\S III]{Jiang02}. It can be seen that
the plaintext is perfectly recovered after a short transient period
of less than one second.

\section{Simulations}

All results were based on simulations with MATLAB 7.1, the Lorenz
integration algorithm was a four-fifth order Runge-Kutta with an
absolute error tolerance of $10^{-9}$, and a relative error
tolerance of $10^{-6}$.

\section{Conclusions}
A simple method was proposed to reduce the parameter search space
of the Lorenz system, based on the determination of the system
equilibrium points from the waveform analysis of one of its
variables $x(t)$. Then the method was applied to the cryptanalysis
of the cryptosystem~\cite{Jiang02}, showing that it is rather weak
since it can be broken without knowing its parameter values. The
total lack of security discourages the use of this algorithm for
secure applications.

\addcontentsline{toc}{section}{Acknowledgment} This work was
supported by Ministerio de Ciencia y Tecnolog\'{\i}a of Spain,
research grant SEG2004-02418, and by The Hong Kong Polytechnic
University's Postdoctoral Fellowships Scheme under grant no.
G-YX63.

\bibliographystyle{IEEEtrans}

\begin{thebibliography}{10}
\providecommand{\url}[1]{#1} \csname url@rmstyle\endcsname
\providecommand{\newblock}{\relax}
\providecommand{\bibinfo}[2]{#2}
\providecommand\BIBentrySTDinterwordspacing{\spaceskip=0pt\relax}
\providecommand\BIBentryALTinterwordstretchfactor{4}
\providecommand\BIBentryALTinterwordspacing{\spaceskip=\fontdimen2\font
plus \BIBentryALTinterwordstretchfactor\fontdimen3\font minus
  \fontdimen4\font\relax}
\providecommand\BIBforeignlanguage[2]{{%
\expandafter\ifx\csname l@#1\endcsname\relax
\typeout{** WARNING: IEEEtran.bst: No hyphenation pattern has been}%
\typeout{** loaded for the language `#1'. Using the pattern for}%
\typeout{** the default language instead.}%
\else \language=\csname l@#1\endcsname \fi #2}}

\bibitem{yang04}
T.~Yang, ``A survey of chaotic secure communication systems,''
\emph{Int. J.
  Comput. Cognit.}, vol.~2, pp. 81--130, June 2004.

\bibitem{Boutayeb02}
M.~Boutayeb, M.~Darouach, and H.~Rafaralahy, ``Generalized
state-space
  observers for chaotic synchronization and secure communication,'' \emph{IEEE
  Trans. Circuits Syst. I-Fundam. Theor. Appl.}, vol.~49, no.~3, pp. 345--349,
  March 2002.

\bibitem{Bowong04}
S.~Bowong, ``Stability analysis for the sinchronization of chaotic
systems with
  different order: application to secure communication,'' \emph{Phys. Lett. A},
  vol. 326, no. 1-2, pp. 102--113, May 2004.

\bibitem{Memon03}
Q.~Memon, ``Synchronized chaos for network security,'' \emph{Comput.
Commun.},
  vol.~26, pp. 498--505, 2003.

\bibitem{Alvarez05b}
G.~{\'A}lvarez, L.~Hern{\'a}ndez, J.~{Mu{\~n}oz}, F.~Montoya, and
S.~Li,
  ``Security analysis of a communication system based on the synchronization of
  different order chaotic systems,'' \emph{Phys. Lett. A}, vol. 345, no.~4, pp.
  245--250, October 2005.

\bibitem{Alvarez04g}
G.~{\'A}lvarez and S.~Li, ``Breaking network security based on
synchronized
  chaos,'' \emph{Comput. Communicat.}, vol.~27, pp. 1679--1681, 2004.

\bibitem{Alvarez04f}
G.~{\'A}lvarez, F.~Montoya, M.~Romera, and G.~Pastor, ``Breaking
two secure
  communication systems based on chaotic masking,'' \emph{IEEE T. Circuits-II},
  vol.~51, no.~10, pp. 505--506, 2004.

\bibitem{Jiang02}
Z.~P. Jiang, ``A note on chaotic secure communication systems,''
\emph{IEEE
  Trans. Circuits Syst. I-Fundam. Theor. Appl.}, vol.~49, no.~1, pp. 92--96,
  2002.

\bibitem{Wang04}
B.-H. Wang and S.~Bu, ``Controlling the ultimate state of projective
  synchronization in chaos: application to chaotic encryption,'' \emph{Int. J.
  Mod. Phys. B}, vol.~18, no. 17--19, pp. 2415--2421, 2004.

\bibitem{Li04}
D.~X. Zhigang~Li, ``A secure communication scheme using projective
chaos synchronization,'' \emph{Chaos, Solitons and Fractals},
vol.~22, pp.
  477--481, 2004.

\bibitem{solak04}
E.~Solak, ``Partial identification of lorenz system and its
application to key   space reduction of chaotic cryptosystems,''
\emph{IEEE Trans. Circuits Syst.
  II-Express briefs}, vol.~51, no.~10, pp. 557--560, October 2004.

\bibitem{Stojanovski96}
T.~Stojanovski, L.~Kocarev, and U.~Parlitz, ``A simple method to
reveal the parameters of the {L}orenz system,'' \emph{J. of
Bifurcation adn Chaos}, vol.~6, no. 12B, pp. 2645--2652, 1996.

\bibitem{Lorenz63}
E.~N. Lorenz, ``Deterministic non periodic flow,'' \emph{J. Atmos.
Sci.},
  vol.~20, no.~2, pp. 130--141, 1963.

\bibitem{Sparrow82}
C.~Sparrow, \emph{The Lorenz equations}, ser. Applied mathematical
  sciences.\hskip 1em plus 0.5em minus 0.4em\relax Springer-Verlag, 1982.

\bibitem{Pecora91}
L.~M. Pecora and T.~L. Carroll, ``Driving systems with chaotic
signals,''
  \emph{Phys. Rev. A}, vol.~44, pp. 2374--2383, 1991.

\end{thebibliography}

\end{document}